# Thick-target transmission method for excitation functions of interaction cross sections

M. Aikawa[1,*], S. Ebata[1], S. Imai[2]

[1] Faculty of Science, Hokkaido University, Sapporo 060-0810, Japan

[2] Institute for the Advancement of Higher Education, Hokkaido University, Sapporo 060-0810, Japan

[*] aikawa@sci.hokudai.ac.jp

**Abstract**

We propose a method, called as thick-target transmission (T3) method, to obtain an excitation function of interaction cross sections. In an ordinal experiment to measure the excitation function of interaction cross sections by the transmission method, we need to change the beam energy for each cross section. In the T3 method, the excitation function is derived from the beam attenuations measured at the targets of different thicknesses without changing the beam energy. The advantage of the T3 method is the simplicity and availability for radioactive beams. To confirm the availability, we perform a simulation for the $^{12}$C+$^{27}$Al system with the PHITS code instead of actual experiments. Our results have large uncertainties but well reproduce the tendency of the experimental data.

**Keywords:** thick-target, transmission method, radioactive beam

**Introduction**

Transmutation of long-lived fission products (LLFP) is one of the key technologies to reduce radioactive wastes produced by nuclear power plants [1]. Nuclear data, such as cross sections, related to the transmutation are essential to develop the technology. Indeed, neutron capture cross sections of LLFP and of minor actinides are measured [2], although experiments of such radioactive isotope (RI) targets have severe restrictions due to high radioactivity and the chemical instability. To avoid such restrictions, experiments in inverse kinematics are available for charged-particle induced data. The RIs including LLFP are actually utilized as a beam in the present accelerators [3] and the cross section data of nuclear wastes, $^{90}$Sr and $^{137}$Cs, have recently been measured [4].

In addition to the cross section data, an essential quantity for the transmutation is thick-target yields (TTY) [5,6,7]. If we require information for the transmutation of LLFP lumps, the TTY plays a key role. It is so hard to measure the TTY with a LLFP target in accelerators due to the target properties. Hence, we suggested a method to estimate the TTY using inverse kinematics instead of LLFP targets [8].

The TTY can be described by the integration of the cross sections with respect to a path length in a target [8,9]. To derive the TTY from cross sections, the excitation function is required. The stacked-foil activation method is available to derive the excitation function for stable targets. The excitation function of the reactions on LLFP targets is however difficult to obtain due to the radioactivity.

Among many kinds of cross sections an interaction cross section is often used in the studies for nuclear size and radii of radioactive isotopes [10,11,12]. It is an inclusive cross section consisting of many channels changing the number of protons or neutrons in a projectile. The interaction cross sections of LLFP on a stable target correspond to the cross sections of the transmutation reaction on the LLFP target. Therefore, a measurement of the excitation function of the interaction cross sections leads to the TTY of LLFP transmutation.

In the ordinal experiments to measure the excitation function of interaction cross sections by the transmission method, a large experimental effort is required to change the beam energy for each cross section. In addition, as summarized in Ref. [12], energies at which cross sections measured are several hundreds MeV/nucleon since RI beams are obtained through spallation reactions of primary beams. In this paper, we propose the thick-target transmission (T3) method to obtain the excitation functions of the interaction cross sections at lower energies. The T3 method consists of iterative measurements of beam attenuations with changing only target thicknesses and without readjusting beam settings. In the T3 method, the target also plays a role of the energy degrader [13] for the projectile. This method is available to measure interaction cross sections of RI beams including LLFP in a wide energy range. We apply the T3 method to the specific reaction as an example, and show the suitability of our method.

**Method**

We consider a beam attenuation by nuclear reactions at a length $x$ (cm) from the surface of a target with a thickness $L$ (cm) as shown in Fig. 1. The numbers of incident particles and unreacted particles at $x$ inside the target are denoted as $N_\mathrm{i}(0)$ and $N_\mathrm{i}(x)$, thus

$$N_\mathrm{i}(0) = N_\mathrm{r}(x) + N_\mathrm{i}(x), \qquad (1)$$

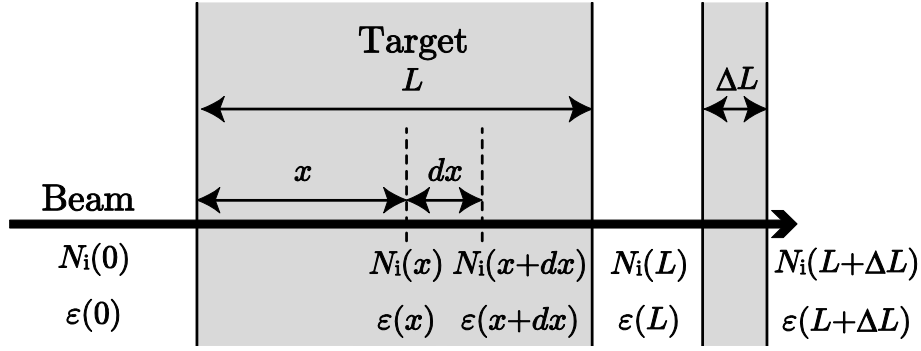

Fig. 1: Schematic of the number of the incident (unreacted) particles $N_i(x)$ and the energy $\varepsilon(x)$ at $x$ from the surface of a target.

where $N_r(x)$ is the number of reacted particles at $x$ (Fig. 1). We assume that the unreacted incident particles can pass through the target while decreasing its energy $\varepsilon(x)$ (MeV/nucleon) by mainly electron scattering. The derivative of Eq. (1) is

$$dN_r(x) + dN_i(x) = 0, \tag{2}$$

since $N_i(0)$ is constant. The $dN_r(x)$ corresponding to the number of reacted particles in $dx$ is deduced from the number density of the target $n_T$ (cm⁻³) and the interaction cross section $\sigma_I(x)$ (cm²) at $x$ as

$$dN_r(x) = N_i(x) n_T \sigma_I(x) dx. \tag{3}$$

From Eqs. (2) and (3),

$$-\frac{1}{N_i(x)} dN_i(x) = n_T \sigma_I(x) dx. \tag{4}$$

We can thus obtain an equation of the beam attenuation from the integral of Eq. (4) as

$$N_i(L) = N_i(0) e^{-Y(L)}, \tag{5}$$

$$Y(L) \equiv n_T \int_0^L \sigma_I(x) dx,$$

where $Y(x)$ is equivalent to the reaction probability per incident particle. Although the form of the attenuation in Eq. (5) is same with that of the decay phenomena, the attenuation is due to the nuclear reaction, and depends on not time but the length. If the target thickness $L$ is sufficiently thin ($L \ll 1$), $Y(L)$ can be approximated as $Y \cong n_T \sigma_I L$ with $\sigma_I$ at the incident energy. As in the ordinary transmission method, the interaction cross section can be derived as

$$\sigma_I = -\frac{1}{n_T L} \ln \frac{N_i(L)}{N_i(0)}. \tag{6}$$

For a thick-target, $Y(L)$ can be derived from Eq. (5) as

$$Y(L) = -\ln\frac{N_i(L)}{N_i(0)}. \tag{7}$$

The $\sigma_I(L)$ value can be derived in an identical manner to the transmission method as

$$\sigma_I(L) = -\frac{1}{n_T}\frac{Y(L+\Delta L) - Y(L)}{\Delta L}$$

$$= -\frac{1}{n_T \Delta L}\ln\left(\frac{N_i(0)}{N_i(L)}\frac{N'_i(L+\Delta L)}{N'_i(0)}\right), \tag{8}$$

which corresponds to the transmission method using the attenuated beam at $L$ in the target with thickness $\Delta L$. $N_i(L)$ and $N'_i(L + \Delta L)$ are obtained in different runs with incident beams of $N_i(0)$ and $N'_i(0)$.

The energy $\varepsilon(L)$ can be measured experimentally and is also estimated using the mass stopping power $S(x) = -\frac{A_P}{\rho}\frac{d\varepsilon(x)}{dx}$ (MeV g$^{-1}$ cm$^2$) [14] as

$$\varepsilon(L) = \varepsilon(0) + \int_0^L \frac{d\varepsilon}{dx}dx$$

$$= \varepsilon(0) - \frac{\rho}{A_P}\int_0^L S(x)dx, \tag{9}$$

where $\rho$ is the density (g cm$^{-3}$) of the target and $A_P$ is the mass number of the projectile.

The excitation function of $\sigma_I(\varepsilon)$ can be obtained from the iterative measurements of $N_i(L)$ with the different thicknesses due to the projectile moderation inside the target. According to the equations, to obtain each $\sigma_I(\varepsilon)$ in a real experiment, we 1) measure the attenuation ratio $N_i(L)/N_i(0)$ and energy $\varepsilon(L)$ in the moderator target with thickness $L$, 2) measure the attenuation ratio $N'_i(L + \Delta L)/N'_i(0)$ and energy $\varepsilon(L + \Delta L)$ in the moderator and reaction targets with thickness $L + \Delta L$, and then 3) obtain $\sigma_I(\varepsilon)$ by Eq. (8) and the median energy of the two cases with target thicknesses $L$ and $L + \Delta L$. The excitation function can be obtained through iterative measurements with different thicknesses, e.g. changing the number of stacked foils.

**Simulation results**

We perform a simulation on the interaction cross sections for the $^{12}$C-induced reaction on $^{27}$Al with the Particle and Heavy Ion Transport code System (PHITS) [15] in order to test the usefulness of the T3 method. Reaction cross sections with a wide energy range for the $^{12}$C+$^{27}$Al system were measured by the conventional transmission method [16,17]. The reaction cross sections exclude the contribution of inelastic scattering from interaction cross sections although the contribution is expected to be small in the present energy region. Our simulation covers the energy range of the experiment and goes lower.

The incident energy of the $^{12}$C beam is set to be 100 MeV/nucleon and the beam stops at around 1.23 cm from the surface of an aluminum target. Accordingly, the maximum thickness of the target is set to be 1.22 cm in the simulation and the unreacted incident particles can pass through the target. The target of the maximum thickness consists of 21 foils with thicknesses of 0.1 cm from 0.0 up to 1.0 cm and of 0.02 cm from 1.0 up to 1.22 cm. The iterative measurements of different thicknesses can be simulated with the stack of the different number of the foils. The trial number is $10^5$ for each target, which corresponds, for example, to the intensity of 1,000 pps and the irradiation period of 100 s. The parameters of the PHITS simulation are shown in Table. 1.

Figure 2 shows the energy distributions of the attenuated beams at $L$ and $L + \Delta L$. The incident $^{12}$C beam with 100 MeV/nucleon is attenuated to 0.91941 at $L = 1.0$ cm with 39.0 MeV/nucleon, and to 0.91758 at $L + \Delta L = 1.02$ cm with 37.1 MeV/nucleon. The distribution of $^{12}$C below the peak is due to nuclear reactions and elastic scattering with the target. In Fig. 3, the attenuation of the $^{12}$C beam (circle) and the energy of unreacted particles (triangle) are summarized as a function of the target thickness. The cross sections $\sigma_I(\varepsilon)$ are obtained from $N_i(L)$ in Eq. (8) and simulated $\varepsilon(L)$ and are shown as a solid line with open circles in Fig. 4. For example, we can obtain $\sigma_I(\varepsilon) = 1.653 \pm 1.056$ (b) at $\varepsilon = 38.0$ (MeV/nucleon) from the described results of $L = 1.0$ and $L + \Delta L = 1.02$. The cross sections reproduce the tendency of experimental data [16,17] in the energy range from 40 to 100 MeV/nucleon. Furthermore, the decreasing behavior of the cross section at the low energy region is found in our simulation, although it has fluctuations and divergence from the expectation values around 30 MeV/nucleon due to poor statistics in the Monte Carlo calculation. If the beam intensity of $10^9$ pps is available, the statistical uncertainty is expected to be reduced by three orders of magnitude.

We emphasize the advantages of the T3 method from the experimental points of view. In cases using beams of RIs, especially far from the stability line, estimations of the PHITS code and other theories may largely differ from those of actual experiments. While maintaining beam conditions, the experiments can be performed by only changing the target thicknesses, e.g., assembling thin foils, which is clearly much easier than the beam adjustments. If a beam consists of several RIs, each intensity of which is ~1,000 pps, the interaction cross sections can be measured simultaneously and systematically for the RIs. The T3 method is essentially restricted only by the beam intensity and statistics. If the LLFP are available as incident particles with enough intensity, the interaction cross sections can be measured and directly related to the discussion of nuclear transmutation. The data can contribute to the study to reduce nuclear wastes.

Table 1: Simulation Parameters

| Projectile | $^{12}$C | |
|---|---|---|
| Energy | 100 MeV/nucleon | |
| Target | $^{27}$Al | |
| Density | 2.7 g/cm$^3$ | |
| Foil thickness | 0.1 cm | ($0 \leq L \leq 1.0$ cm) |
| | 0.02 cm | ($1.0 \leq L \leq 1.22$ cm) |
| Trial number | $10^5$ | |

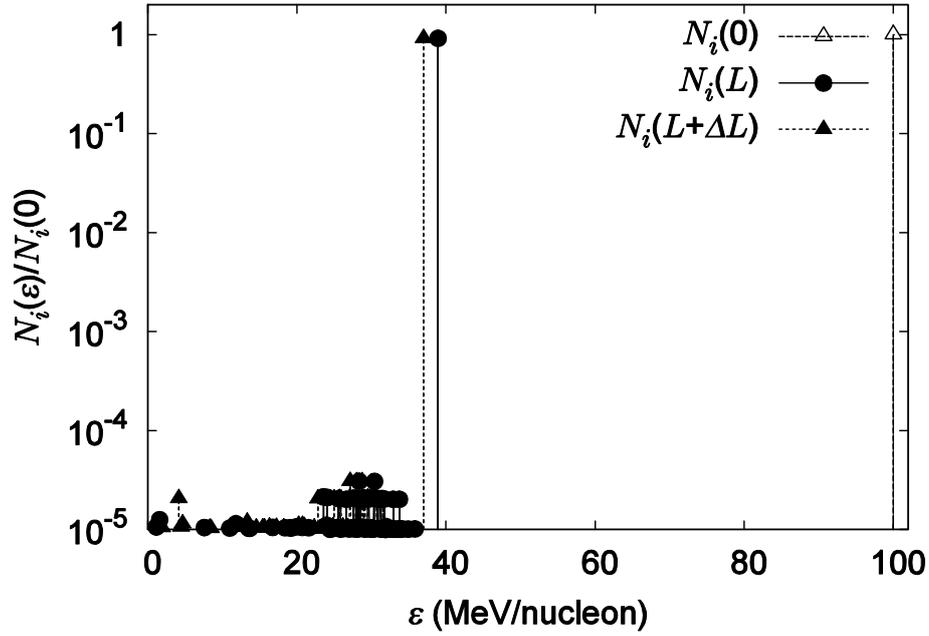

Fig. 2: The energy distribution of $^{12}$C at the upstream and downstream of the targets. The incident beam with 100 MeV/nucleon (open triangle) is attenuated from 1.0 to 0.91941 (solid circles) with 39.0 MeV/nucleon and 0.91758 (solid triangles) with 37.1 MeV/nucleon which are corresponding to at $L = 1.0$ cm and $L + \Delta L = 1.02$ cm, respectively.

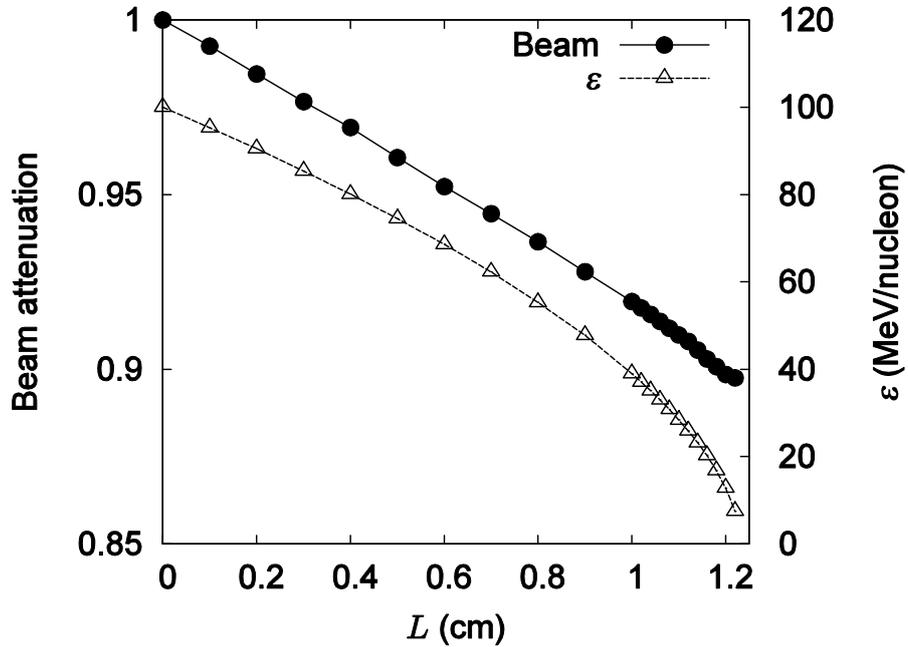

Fig. 3: The attenuation (solid line with solid circles) and energy (dotted line with open triangles) of the incident beam are shown as a function of the thickness of the target.

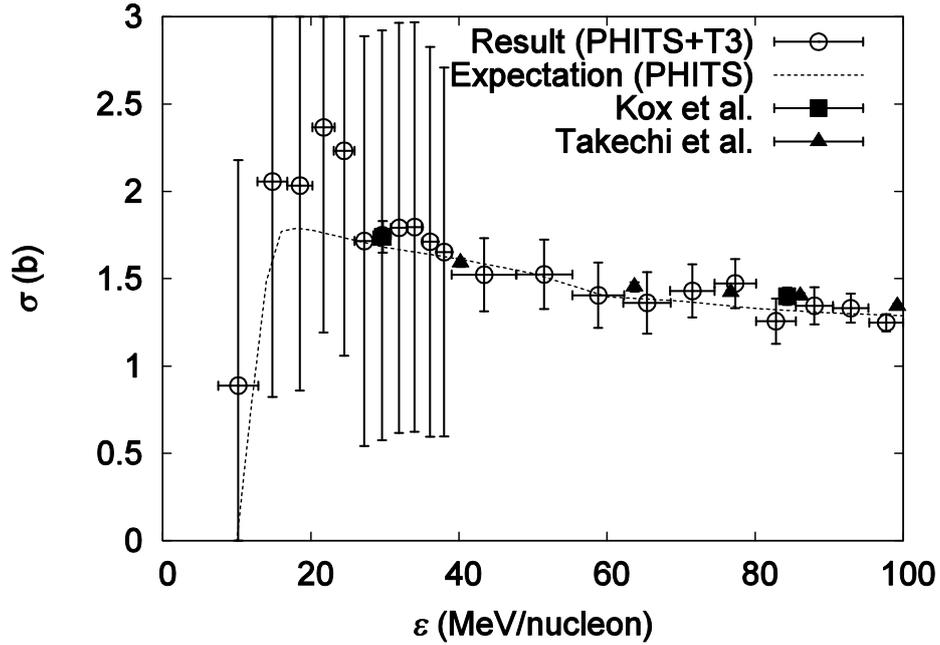

Fig. 4: Interaction or reaction cross section of $^{12}$C on $^{27}$Al. The results of the PHITS simulation (open circles) are compared with the reaction cross section data (solid squares and triangles) [16,17]. The interaction cross section expected in PHITS is also shown as a dotted-line.

**Summary**

We propose the T3 method to measure the excitation functions of interaction cross sections. The T3 method can derive the interaction cross sections through the iterative measurements of beam attenuations with targets of different thicknesses. The method is applied to the system of the $^{12}$C-induced reaction on $^{27}$Al. The beam attenuation ratio $N_i(L)/N_i(0)$ at the target thickness $L$ is simulated by PHITS. The interaction cross section $\sigma_I(L)$ can indeed be derived from the obtained attenuation ratios and Eq. (8) and the result is in a good agreement with the experimental data.

The T3 method is also available for RI beams, the interaction cross sections of which are sometimes difficult to reproduce theoretically. In our simulation based on the realistic condition of RI beams in recent accelerators, the trial number at each thickness is set as $10^5$, which is consistent with an intensity of 1,000 pps and an irradiation time of 100 s. According to our results, we show availability to derive the excitation function using RI beams. The T3 method can contribute to accumulate fundamental data of the reaction yields on radiative matter.

The information led from the interaction cross sections plays a key role to develop the transmutation technology.

**Acknowledgement**

This work was funded by ImPACT Program of Council for Science, Technology and Innovation (Cabinet Office, Government of Japan). The authors are very grateful for valuable comments on the PHITS simulation from Dr. K. Niita (RIST).